\documentclass[10pt]{amsart}
\usepackage{fullpage,graphicx,psfrag,epsfig,amsmath,amsfonts,amsthm,amssymb,fancybox,ulem,url}
\usepackage[dvips]{color}
\begin{document}

\title{Entropy for mixtures of discrete and continuous
variables}


\author{Chandra Nair}
\address{Microsoft Research, Redmond, WA 98052 }
\email{cnair@microsoft.com}

\author{Balaji Prabhakar}
\address{Depts. of EE and CS, Stanford University, Stanford, CA }
\email{balaji@ee.stanford.edu}

\author{Devavrat Shah}
\address{Dept. of EECS, MIT, Cambridge, MA}
\email{devavarat@mit.edu}

\maketitle

\newcommand{\eq}{\begin{equation}}
\newcommand{\en}{\end{equation}}
\newcommand{\eqa}{\begin{eqnarray}}
\newcommand{\ena}{\end{eqnarray}}
\newcommand{\eqas}{\begin{eqnarray*}}
\newcommand{\enas}{\end{eqnarray*}}
\newcommand{\ra}{ {\rightarrow} }
\newcommand{\Ra}{ {\Rightarrow} }
\newcommand{\bX}{\mathbf{X}}
\newcommand{\bY}{\mathbf{Y}}
\newcommand{\bZ}{\mathbf{Z}}
\newcommand{\bU}{\mathbf{U}}
\newcommand{\bV}{\mathbf{V}}
\newcommand{\bW}{\mathbf{W}}
\newcommand{\p}{{\rm P}}
\def\mb#1{\mathbf{#1}}
\newenvironment{myproof}[1]{\noindent\hspace{2em}{\itshape Proof #1:}}{\hspace*{\fill}~\QED\par\endtrivlist\unskip}

\def\cA{{\mathcal A}}
\def\cB{{\mathcal B}}
\def\cC{{\mathcal C}}
\def\cD{{\mathcal D}}
\def\cE{{\mathcal E}}
\def\cF{{\mathcal F}}
\def\cG{{\mathcal G}}
\def\cH{{\mathcal H}}
\def\cI{{\mathcal I}}
\def\cJ{{\mathcal J}}
\def\cK{{\mathcal K}}
\def\cL{{\mathcal L}}
\def\cM{{\mathcal M}}
\def\cN{{\mathcal N}}
\def\cO{{\mathcal O}}
\def\cP{{\mathcal P}}
\def\cQ{{\mathcal Q}}
\def\cR{{\mathcal R}}
\def\cS{{\mathcal S}}
\def\cT{{\mathcal T}}
\def\cU{{\mathcal U}}
\def\cV{{\mathcal V}}
\def\cW{{\mathcal W}}
\def\cX{{\mathcal X}}
\def\cY{{\mathcal Y}}
\def\cZ{{\mathcal Z}}


\def\tw{{\textbf w}}
\def\ty{{\textbf y}}
\def\tx{{\textbf x}}
\def\ta{{\textbf a}}
\def\tbr{{\textbf r}}
\def\tbb{{\textbf b}}
\def\tC{{\textbf C}}
\def\TV{{\textbf V}}

\def\bbN{{\mathbb N}}
\def\bbR{{\mathbb R}}
\def\bbP{{\mathbb P}}
\def\bbS{{\mathbb S}}
\def\bbZ{{\mathbb Z}}


\definecolor{Red}{rgb}{1,0,0}
\definecolor{Blue}{rgb}{0,0,1}
\definecolor{Green}{rgb}{0,1,0}
\definecolor{Yellow}{rgb}{1,1,0}
\definecolor{Cyan}{rgb}{0,1,1}
\definecolor{Magenta}{rgb}{1,0,1}
\definecolor{Orange}{rgb}{1,.5,0}
\definecolor{Violet}{rgb}{.5,0,.5}
\definecolor{Purple}{rgb}{.75,0,.25}
\definecolor{Brown}{rgb}{.75,.5,.25}
\definecolor{Grey}{rgb}{.5,.5,.5}

\def\red{\color{Red}}
\def\blue{\color{Blue}}
\def\green{\color{Green}}
\def\yellow{\color{Yellow}}
\def\cyan{\color{Cyan}}
\def\magenta{\color{Magenta}}
\def\orange{\color{Orange}}
\def\violet{\color{Violet}}
\def\purple{\color{Purple}}
\def\brown{\color{Brown}}
\def\grey{\color{Grey}}

\def\tg{\tilde{g}}
\def\tr{\tilde{r}}
\def\tb{\tilde{b}}
\def\tU{\tilde{U}}
\def\tV{\tilde{V}}
\def\tX{\tilde{X}}
\def\tW{\tilde{W}}
\def\tY{\tilde{Y}}
\def\tZ{\tilde{Z}}

\def\hZ{\hat{Z}}

\def\bij{\rightleftharpoons}

\def\H{\mathbb H}
\def\I{\mathbb I}

\def\cc{\bar{\mbox{co}}}

\newtheorem{theorem}{Theorem}[section]
\newtheorem{lemma}[theorem]{Lemma}
\newtheorem{claim}[theorem]{Claim}
\newtheorem{conjecture}[theorem]{Conjecture}
\newtheorem{corollary}[theorem]{Corollary}

\newtheorem{proposition}[theorem]{Proposition}

\theoremstyle{definition}
\newtheorem{definition}{Definition}[section]
\newtheorem{example}[definition]{Example}
\newtheorem{xca}[definition]{Exercise}

\theoremstyle{remark}
\newtheorem{remark}{Remark}[section]

\newtheorem{fact}[remark]{Fact}

\newcommand{\enp} {\hfill \rule{2.2mm}{2.6mm}}
\def\mb#1{\mathbf{#1}}
\numberwithin{equation}{section}


\renewcommand{\thefootnote}{\fnsymbol{footnote}}



\newcommand{\blankbox}[2]{%
  \parbox{\columnwidth}{\centering
    \setlength{\fboxsep}{0pt}%
    \fbox{\raisebox{0pt}[#2]{\hspace{#1}}}%
  }%
}

\begin{abstract}

In this paper, we extend the notion of entropy in a natural manner
for a mixed-pair random variable, a pair of random variables with one
discrete and the other continuous. Our extensions are consistent
in that there exist natural injections from discrete or continuous
random variables into mixed-pair random variables such that their
entropy remains the same. This extension of entropy allows us to
obtain sufficient conditions for the entropy preservation under
bijections between mixed-pair random variables.

The extended definition of entropy leads to an entropy rate for
continuous time Markov chains. As applications of our results, we
provide simpler proofs of some known probabilistic results. The
frame-work developed in this paper is best suited for establishing
probabilistic properties of complex processes, such as load
balancing systems, queuing networks, caching algorithms, that have
inherent discrete variables (choices made) and continuous
variables (occurrence times).

\end{abstract}



\section{Introduction}

The notion of entropy for discrete random variables as well as
continuous random variables is well defined. Entropy preservation
of discrete random variable under bijection map is an extremely
useful property. For example, Prabhakar and Gallager \cite{prg03}
used this entropy preservation property to obtain an alternate
proof of the known result that Geometric processes are fixed
points under certain queuing disciplines.

In many interesting situations, including Example
\ref{ex:poi-split} given below, the underlying random variables
are mixtures of discrete and continuous random variables. Such
systems exhibit natural bijective properties which allow one to
obtain non-trivial properties of the system via {\it non-rigorous}
`` information preservation" arguments. In this paper we develop
sufficient conditions to make such arguments rigorous.

We will extend the definition of entropy to random variables that
form a mixed pair of discrete and continuous variables as well as
obtain sufficient conditions for preservation of entropy.
Subsequently, we will provide a rigorous justification of
mathematical identities that follow in the example below.

\medskip

\begin{example}
\label{ex:poi-split} {\it Poisson Splitting:}  Consider a Poisson
Process, $\cP$, of rate $\lambda$. Split the Poisson process into
two baby-processes $\cP_1$ and $\cP_2$ as follows: for each point
of $\cP$, toss an independent coin of bias $p$; if coin turns up
heads then the point is assigned to $\cP_1$, else to $\cP_2$. It
is well-known that $\cP_1$ and $\cP_2$ are independent Poisson
processes with rates $\lambda p$ and $\lambda(1-p)$ respectively.

Entropy rate of a Poisson process with rate $\mu$ is known to be
$\mu (1- \log \mu)$ nats per second. That is, entropy rates of
$\cP$, $\cP_1$, and  $\cP_2$ are given by $\lambda(1- \log
\lambda)$, $\lambda p(1- \log \lambda p)$ and $\lambda (1-p)(1-
\log \lambda(1-p))$ respectively. Further observe that the coin of
bias $p$ is tossed at a rate $\lambda$ and each coin-toss has an
entropy equal to $- p \log p - (1-p) \log (1-p)$ nats.

It is clear that there is a bijection between the tuple ($\cP$,
coin-toss process) and the tuple ($\cP_1$,$\cP_2$).  Observe that
the joint entropy rate of the two independent baby-processes are
given by their sum. This leads to the following ``obvious'' set of
equalities.

\begin{equation}
\label{eq:poisson}
\begin{split} &H_{ER}(\cP_1,\cP_2) = H_{ER}(\cP_1) + H_{ER}(\cP_2) \\
&\quad =  \lambda p
(1-\log \lambda p) + \lambda (1-p)(1- \log \lambda(1-p)) \\
&\quad = \lambda (1-\log \lambda) + \lambda(- p  \log p - (1-p) \log (1-p)) \\
&\quad = H_{ER}(\cP) + \lambda(- p  \log p - (1-p) \log (1-p)).
\end{split}
\end{equation}
\normalsize

The last sum can be identified as sum of the entropy rate of the
original Poisson process and the entropy rate of the coin tosses.
However the presence of differential entropy as well as discrete
entropy prevents this interpretation from being rigorous. In this
paper, we shall provide rigorous justification to the above
equalities.

\end{example}

\section{Definitions and Setup}\label{se:sec1}

This section provides technical definitions and sets up the
frame-work for this paper. First, we present some preliminaries.

\subsection{Preliminaries}

Consider a measure space $(\Omega, \cF, \bbP)$, with $\bbP$ being
a probability measure. Let $(\bbR, B_\bbR)$ denote the measurable
space on $\bbR$ with the Borel $\sigma$-algebra. A random variable
$X$ is a measurable mapping from $\Omega$ to $\bbR$. Let $\mu_X$
denote the induced probability measure on $(\bbR, B_\bbR)$ by $X$.
We call $X$ as {\it discrete random variable} if there is a
countable subset $\{x_1,x_2,...\}$ of $\bbR$ that forms a support
for the measure $\mu_X$. Let $p_i = \bbP(X=x_i)$ and note that
$\sum_i p_i = 1$.

The entropy of a discrete random variable is defined by the sum
\[ H(X) = -\sum_i p_i \log p_i. \] Note that this entropy is
non-negative and has several well known properties. One natural
interpretation of this number is in terms of the maximum
compressibility (in bits per symbol) of an i.i.d. sequence of the
random variables, $X$ (cf. Shannon's data compression theorem
\cite{sha48}).

A random variable $Y$, defined on $(\Omega, \cF, \bbP)$,  is said
to be a {\it continuous random variable} if the probability
measure, $\mu_Y$, induced on $(\bbR, B_{\bbR})$ is absolutely
continuous with respect to the Lebesgue measure. These probability
measures can be characterized by a non-negative density function
$f(x)$ that satisfies $\int_{\bbR} f(x)dx = 1$. The entropy
(differential entropy) of a continuous random variable is defined
by the integral
\[ h(Y) = - \int_{\bbR} f(y) \log f(y) dy. \]

The entropy of a continuous random variable is not non-negative,
though it satisfies several of the other properties of the
discrete entropy function. Due to negativity, differential entropy
clearly does  not have interpretation of maximal compressibility.
However, it does have the interpretation of being the limiting
difference between the maximally compressed quantization of the
random variable and an identical quantization of an independent
U$[0,1]$\footnote{ U[0,1] represents a random variable that is
uniformly distributed on the interval [0,1]} random variable
\cite{cot91}  as the quantization resolution goes to zero. Hence
the term differential entropy is usually preferred to entropy when
describing this number.

\subsection{Our Setup}

In this paper, we are interested in a set of random variables that
incorporate the aspects of both discrete and continuous random
variables. Let $Z=(X,Y)$ be a measurable mapping from the space
$(\Omega, \cF, \bbP)$ to the space $(\bbR \times \bbR, B_\bbR
\times B_\bbR)$. Observe that this mapping induces a probability
measure $\mu_Z$ on the space $(\bbR \times \bbR, B_\bbR \times
B_\bbR)$ as well as two probability measures $\mu_X$ and $\mu_Y$
on $(\bbR, B_{\bbR})$ obtained via the projection of the measure
$\mu_Z$.

\begin{definition}[Mixed-Pair]
Consider a random variables $Z=(X,Y)$. We call $Z$\footnote{For
the rest of the paper we shall adopt the notation that random
variables $X_i$ represent discrete random variables, $Y_i$
represent continuous random variables and $Z_i$ represent
mixed-pair of random variables.} a mixed-pair if $X$ is a discrete
random variable while $Y$ is a continuous random variable. That
is, the support of $\mu_Z$ is on the product space $\bbS \times
\bbR$, with $\bbS = \{x_1,x_2,...\}$ is a countable subset of
$\bbR$. That is $\bbS$ forms a support for $\mu_X$ while $\mu_Y$
is absolutely continuous with respect to the Lebesgue measure.
\end{definition}

Observe that 
$Z = (X, Y)$ induces measures $\{\mu_1, \mu_2,.... \}$ that are
absolutely continuous with respect to the Lebesgue measure, where
$\mu_i(A) = \bbP(X=x_i,Y\in A)$, for every $A \in B_\bbR$.
Associated with these  measures $\mu_i$, there are non-negative
density functions $g_i(y)$ that satisfy
\[ \sum_i \int_\bbR g_i(y)dy = 1.\]

Let us define $p_i = \int_\bbR g_i(y)dy$. Observe that $p_i$'s are
non-negative numbers that satisfy $\sum_i p_i = 1$ and corresponds
to the probability measure $\mu_X$. Further $g(y) = \sum_i g_i(y)$
corresponds to the probability measure $\mu_Y$. Let
$$\tg_i(y) \stackrel{\triangle}{=}\frac{1}{p_i} g_i(y)$$ be the
probability density function of $Y$ conditioned on $X=x_i$.

The following non-negative sequence is well defined for every $y
\in \bbR$ for which $g(y)
> 0$,
\[ p_i(y) = \frac{g_i(y)}{g(y)}, ~i\geq 1.\]
Now $g(y)$ is finite except possibly on a set, $A$, of measure
zero. For $y \in A^c$, we have that $\sum_i p_i(y) = 1$; $p_i(y)$
corresponds to the probability that $X=x_i$ conditioned on $Y=y$.
It follows from definitions of $p_i$ and $p_i(y)$ that
\[ p_i = \int_\bbR p_i(y) g(y) dy. \]

\medskip

\begin{definition}[Good Mixed-Pair ] A mixed-pair random
variable $Z=(X,Y)$ is called {\it good} if the following condition
is satisfied:
\begin{equation}
\label{eq:goodmixedpair} \sum_i \int_\bbR |g_i(y) \log g_i(y)| dy
< \infty.
\end{equation}
\end{definition}

\medskip

Essentially, the good mixed-pair random variables possess the
property that when restricted to any of the $X$ values, the
conditional differential entropy of $Y$ is well-defined.  The
following lemma provides a simple sufficient conditions for
ensuring that a mixed-pair variable is good.

\begin{lemma}
\label{le:gmpcond} The following conditions are sufficient for a
mixed-pair random variable to be a good pair:
\begin{itemize}
\item[(a)] Random variable $Y$ possess a finite $\epsilon^{th}$
moment for some $\epsilon > 0$, i.e.
$$ M_\epsilon = \int_\bbR |y|^\epsilon g(y) dy < \infty. $$
\item[(b)]There exists $\delta > 0$ such that $g(y)$
satisfies
$$ \int_{\bbR} g(y)^{1+\delta} dy < \infty.$$
\item[(c)] The discrete random variable $X$ has finite entropy,
i.e. $- \sum_i p_i \log p_i < \infty$.
\end{itemize}
\end{lemma}

\medskip

\begin{proof}
The proof is presented in the appendix.
\end{proof}

\medskip

\begin{definition}[Entropy of a mixed-pair]
The entropy of a good mixed-pair random variable is defined by
\begin{equation}
\label{eq:ent-cross} \H(Z) = - \sum_i \int_\bbR g_i(y) \log g_i(y)
dy.
\end{equation}
\end{definition}

\medskip

\begin{definition}[Vector of Mixed-Pairs]
Consider a random vector $(Z_1,...,Z_d)=\{(X_1,Y_1), ... ,
(X_d,Y_d)\}$. We call $(Z_1,...,Z_d)$ a vector of mixed-pairs if
the support of $\mu_{(Z_1,...,Z_d)}$ is on the product space
$\bbS^d \times \bbR^d$, where $\bbS^d \subset \bbR^d$ is a
countable set. That is, $\bbS^d$ forms the support for the
probability measure $\mu_{(X_1,..,X_d)}$ while the measure
$\mu_{(Y_1,..,Y_d)}$  is absolutely continuous with respect to the
Lebesgue measure on $\bbR^d$.
\end{definition}

\begin{definition}[Good Mixed-Pair Vector ] A vector of mixed-pair random
variables $(Z_1,...,Z_d)$ is called good if the following
condition is satisfied:
\begin{equation}
\label{eq:goodmixedpairvec} \sum_{\tx \in \bbS^d} \int_{\ty \in
\bbR^d} |g_\tx(\ty) \log g_\tx(\ty)| d\ty < \infty,
\end{equation}
where $g_\tx(\ty)$ is the density of the continuous random vector
$Y^d$ conditioned on the event that $X^d = \tx$.
\end{definition}

Analogous to Lemma \ref{le:gmpcond}, the following conditions
guarantee that a vector of mixed-pair random variables is good.

\begin{lemma}
\label{le:gmpcondvec} The following conditions are sufficient for
a mixed-pair random variable to be a good pair:
\begin{itemize}
\item[(a)] Random variable $Y^d$ possess a finite $\epsilon^{th}$
moment for some $\epsilon > 0$, i.e.
$$ M_\epsilon = \int_{\bbR^d} \|\ty\|^\epsilon g(\ty) d\ty < \infty. $$
\item[(b)]There exists $\delta > 0$ such that $g(\ty)$
satisfies
$$ \int_{\bbR^d} g(\ty)^{1+\delta} d\ty < \infty.$$
\item[(c)] The discrete random variable $X^d$ has finite entropy,
i.e. $- \sum_{\tx \in \bbS^d} p_\tx \log p_\tx < \infty$.
\end{itemize}
\end{lemma}
\begin{proof}
The proof is similar to that of Lemma \ref{le:gmpcond} and is
omitted.
\end{proof}

 In rest of the paper, all mixed-pair variables and vectors are assumed to be
{\it good}, i.e. assumed to satisfy the condition
\eqref{eq:goodmixedpair}.

\medskip

\begin{definition}[Entropy of a mixed-pair vector]
The entropy of a good mixed-pair vector of random variables is
defined by
\begin{equation}
\label{eq:ent-crossvec} \H(Z) = - \sum_{\tx \in \bbS^d}
\int_{\bbR^d} g_\tx(\ty) \log g_\tx(\ty) d\ty.
\end{equation}
\end{definition}

\medskip

\begin{definition}[Conditional entropy]
Given a pair of random variables $(Z_1,Z_2)$, the conditional
entropy is defined as follows
$$ \H(Z_1|Z_2) = \H(Z_1,Z_2) - \H(Z_2). $$

It is not hard to see that $\H(Z_1|Z_2)$ evaluates to
\begin{equation*} - \sum_{x_1,x_2} \int_{\bbR^2} g_{x_1,x_2}(y_1,y_2) \log
\frac{g_{x_1,x_2}(y_1,y_2)}{g_{x_2}(y_2)} dy_1 dy_2.
\end{equation*}

\end{definition}

\medskip

\begin{definition}[Mutual Information]
Given a pair of random variables $(Z_1,Z_2)$, the mutual
information is defined as follows
$$ \I(Z_1;Z_2) = \H(Z_1) + \H(Z_2) - \H(Z_1,Z_2). $$
\end{definition}
\medskip

The mutual information evaluates to
\begin{equation*}
\sum_{x_1,x_2} \int_{\bbR^2} g_{x_1,x_2}(y_1,y_2) \log
\frac{g_{x_1,x_2}(y_1,y_2)}{g_{x_1}(y_1)g_{x_2}(y_2)} dy_1 dy_2.
\end{equation*}

Using the fact that $ 1 + \log x < x$ for $x > 0$ it can be shown
that $\I(Z_1;Z_2)$ is non-negative.

\medskip

\subsection{Old Definitions Still Work}

We will now present injections from the space of discrete (or
continuous) random variables into the space of mixed-pair random
variable so that the entropy of the mixed-pair random variable is
the same as the discrete (or continuous) entropy.

\vspace{.1in} \noindent {\it Injection: Discrete into Mixed-Pair}:
Let $X$ be a discrete random variable with finite entropy. Let
$\{p_1,p_2,\ldots\}$ denote the probability measure associated
with $X$.  Consider the mapping $\sigma_d: X \to Z \equiv (X,U)$
where $U$ is an independent continuous random variable distributed
uniformly on the interval [0,1]. For $Z$, we have $g_i(y) = p_i$
for $y \in [0,1]$. Therefore
\begin{equation*}
\begin{split}
\H(Z) & =  - \sum_i \int_\bbR g_i(y) \log g_i(y) \; dy  = \sum_i \int_0^1 - p_i \log p_i dy \\
 & = - \sum_i p_i \log p_i  = H(X) < \infty.
\end{split}
\end{equation*}
Therefore we see that $\H(Z) = H(X)$.

\vspace{.1in} \noindent {\it Injection: Continuous into
Mixed-Pair}: Let $Y$ be a continuous random variable with a
density function $g(y)$ that satisfies
\[ \int_\bbR g(y) |\log g(y)| \; dy < \infty. \]
Consider the mapping $\sigma_c: Y \to Z \equiv (X_0,Y)$ where
$X_0$ is the constant random variable, say  $\bbP(X_0 = 1) = 1$.
Observe that $g(y) = g_1(y)$ and that the pair $Z \equiv (X_0,Y)$
is a good mixed-pair that satisfies $\H(Z) = h(Y)$.

Thus $\sigma_d$ and $\sigma_c$ are injections from the space of
continuous and discrete random variables into the space of good
mixed-pairs that preserve the entropy function.

\subsection{Discrete-Continuous Variable as Mixed-Pair}

Consider a random variable\footnote{Normally such random variables
are referred to as mixed random variables.} $V$ whose support is
combination of both discrete and continuous. That is, it satisfies
the following properties: (i) There is a countable set (possibly
finite) $S = \{x_1,x_2,... \}$ such that $\mu_V(x_i)=p_i > 0$;
(ii) measure $\tilde{\mu}_V$ with an associated non-negative
function $\tg(y)$ (absolutely continuous w.r.t. the Lebesgue
measure), and (iii) the following holds:
\[ \int_{\bbR} \tg(y) \; dy + \sum_i p_i = 1. \]
Thus, the random variable $V$ either takes discrete values $x_1,
x_2, \ldots $ with probabilities $p_1,p_2, \ldots$ or else it is
distributed according to the density function $\frac{1}{1-p}
\tg(y)$; where $p = \sum_i p_i$. Observe that $V$ has neither a
countable support nor is its measure absolutely continuous with
respect to Lebesgue measure. Therefore, though such random
variables are encountered neither the discrete entropy  nor the
continuous entropy is appropriate.

To overcome this difficulty, we will treat such variables as
mixed-pair variables by appropriate injection of such variables
into mixed-pair variables. Subsequently, we will be able to use
the definition of entropy for mixed-pair variables.

\vspace{.1in} \noindent{\it Injection: Discrete-Continuous into
Mixed-Pair}: Let $V$ be a discrete-continuous variable as
considered above. Let the  following two conditions be satisfied:
\[ -\sum_i p_i \log p_i < \infty ~ \mbox{and} \int_{\bbR} \tg(y) |\log
\tg(y)| \; dy < \infty. \] Consider the mapping $\sigma_m: V \to Z
\equiv (X,Y)$ described as follows: When $V$ takes a discrete
value $x_i$, it is mapped on to the pair $(x_i, u_i)$ where $u_i$
is chosen independently and uniformly at random in  $[0,1]$. When
$V$ does not take a discrete value and say takes value $y$, it
gets mapped to the pair $(x_0,y)$ where $x_0 \neq x_i, \forall i.$
One can think of $x_0$ as an indicator value that $V$ takes when
it is not discrete. The mixed-pair variable $Z$ has its associated
functions $\{g_0(y), g_1(y), ... \}$ where $g_i(y) = p_i,
~y\in[0,1], i \geq 1$ and $g_0(y) = \tg(y)$.  The entropy of $Z$
as defined earlier is
\begin{equation*}
\begin{split}
&\H(Z)  = - \sum_i \int_\bbR g_i(y) \log g_i(y) \; dy \\
&\quad ~\quad = - \sum_i p_i \log p_i - \int_\bbR \tg(y) \log
\tg(y) \; dy.
\end{split}
\end{equation*}

\begin{remark}
In the rest of the paper we will treat every random variable that
is encountered as a mixed-pair random variable. That is, a
discrete variable or a continuous variable  would be assumed to be
injected into the space of mixed-pairs using the map $\sigma_d$ or
$\sigma_c$, respectively.
\end{remark}

\section{Bijections and Entropy Preservation}
\label{se:sec2}

In this section we will consider bijections between mixed-pair
random variables and establish sufficient conditions under which
the entropy is preserved. We first consider the case of mixed-pair
random variables and then extend this to vectors of mixed-pair
random variables.

\subsection{Bijections between Mixed-Pairs}

Consider mixed-pair random variables $Z_1 \equiv (X_1,Y_1)$ and
$Z_2 \equiv (X_2,Y_2) $. Specifically, let $\bbS_1 = \{x_{1i} \}$
and $\bbS_2 =\{x_{2j} \}$ be the countable (possibly finite)
supports of the discrete measures $\mu_{X_1}$ and $\mu_{X_2}$ such
that $\mu_{X_1}(x_{1i}) > 0 $ and $\mu_{X_2}(x_{2j}) > 0 $ for all
$i \in \bbS_1$ and $j \in \bbS_2$. Therefore a bijection between
mixed-pair variables $Z_1$ and $Z_2$ can be viewed as bijections
between $\bbS_1 \times \bbR$ and $\bbS_2 \times \bbR$.

Let $F : \bbS_1 \times \bbR \to \bbS_2 \times \bbR$ be a
bijection. Given $Z_1$, this bijection induces a mixed-pair random
variable $Z_2$. We restrict our attention to the case when $F$ is
continuous and differentiable\footnote{The continuity of mapping
between two copies of product space $\bbS \times \bbR$ essentially
means that the mapping is continuous with respect to right (or
$Y$) co-ordinate for fixed $x_i \in \bbS$. Similarly,
differentiability essentially means differentiability with respect
to $Y$ co-ordinate.}. Let the induced projections be $F_d: \bbS_1
\times \bbR \to \bbS_2$ and $F_c: \bbS_1 \times \bbR \to \bbR$.
Let the associated projections of the inverse map $F^{-1}: \bbS_2
\times \bbR \to \bbS_1 \times \bbR$ be  $F^{-1}_d: \bbS_2 \times
\bbR \to \bbS_1$ and $F_c^{-1}: \bbS_2 \times \bbR \to \bbR$
respectively.

As before, let $\{g_i(y_1)\}$, $\{h_j(y_2)\}$ denote the
non-negative density functions associated with the mixed-pair
random variables $Z_1$ and $Z_2$ respectively. Let  $(x_{2j},y_2)
= F(x_{1i},y_1)$, i.e. $x_{2j} = F_d(x_{1i},y_1)$ and $y_2 =
F_c(x_{1i},y_1)$.  Now, consider a small neighborhood $x_{1i}
\times [y_1, y_1 +dy_1)$ of $(x_{1i}, y_1)$. From the continuity
of $F$, for small enough $dy_1$, the neighborhood $x_{1i} \times
[y_1, y_1 +dy_1)$  is mapped to some small neighborhood of
$(x_{2j}, y_2)$, say $x_{2j} \times [y_2, y_2+dy_2)$. The measure
of $x_{1i} \times [y_1, y_1 +dy_1)$ is $\approx g_i(y_1) |dy_1|$,
while measure of $x_{2j} \times [y_2, y_2+dy_2)$ is $\approx
h_j(y_2) |dy_2|$. Since distribution of $Z_2$ is induced by the
bijection from $Z_1$, we obtain
\begin{equation}
\label{eq:chofmeasure}
\begin{split}
& g_i(y_1) \left|\frac{dy_1}{dy_2} \right|= h_j(y_2). \\
\end{split}
\end{equation}
Further from $y_2 = F_c(x_{1i},y_1)$ we also have,
\begin{equation}
\label{eq:chofmeasure1}
\begin{split}
& \frac{dy_2}{dy_1} = \frac{dF_c(x_{1i},y_1)}{d y_1}. \\
\end{split}
\end{equation}
These immediately imply a sufficient condition under which
bijections between mixed-pair random variables imply that their
entropies are preserved.
\begin{lemma}
\label{le:bijpre} If $\left|\frac{dF_c(x_{1i},y_1)}{d y_1}\right|
= 1$ for all points $(x_{1i},y_1) \in \bbS_1 \times \bbR$, then
$\H(Z_1) = \H(Z_2)$.
\end{lemma}
\begin{proof}
This essentially follows from the change of variables and repeated
use of Fubini's theorem (to interchange the sums and the
integral). To apply Fubini's theorem, we use the assumption that
mixed-pair random variables are {\it good}. Observe that,
\begin{equation}
\label{eq:bijpreserve}
\begin{split}
 \H(Z_1)  &= - \sum_i \int_{\bbR} g_i(y_1) \log g_i(y_1) dy_1 \\
&  \stackrel{(a)}{=} - \sum_j \int_{\bbR} h_j(y_2) \log
\left( h_j(y_2) \left|\frac{dF_c(x_{1i},y_1)}{dy_1}\right| \right) dy_2 \\
&  \stackrel{(b)}{=} - \sum_j \int_{\bbR} h_j(y_2) \log h_j(y_2) dy_2 \\
& = \H(Z_2).
\end{split}
\end{equation}
Here $(a)$ is obtained by repeated use of Fubini's theorem along
with \eqref{eq:chofmeasure} and $(b)$ follows from the assumption
of the Lemma that $|\frac{dF_c(x_{1i},y_1]}{dy_1}| = 1$.
\end{proof}

\subsection{Some Examples}

In this section, we present some examples to illustrate our
definitions, setup and the entropy preservation Lemma.
\label{sse:example}
\begin{example}
\label{ex:disc}  Let $Y_1$ be a continuous random variable that is
uniformly distributed in the interval $[0,2]$. Let $X_2$ be the
discrete random variable that takes value $0$ when $Y_1 \in [0,1]$
and $1$ otherwise. Let $Y_2 = Y_1 - X_2$. Clearly $Y_2 \in [0,1]$,
is uniformly distributed and independent of $X_2$.

Let $Z_1 \equiv (X_1,Y_1)$ be the natural injection, $\sigma_c$ of
$Y_1$ (i.e. $X_1$ is just the constant random variable.). Observe
that the bijection between $Z_1$ to the pair $Z_2 \equiv
(X_2,Y_2)$ that satisfies conditions of Lemma \ref{le:bijpre} and
implies
\[ \log 2 = \H(Z_1) = \H(Z_2).  \]

However, also observe that by plugging in the various definitions
of entropy in the appropriate spaces, $\H(Z_2) = \H(X_2,Y_2) =
H(X_2) + h(Y_2) = \log 2 + 0$, where the first term is the
discrete entropy and the second term is the continuous entropy. In
general it is not difficult to see that the two definitions of
entropy (for discrete and continuous random variables) are
compatible with each other if the random variables themselves are
thought of as a mixed-pair.

\end{example}

\begin{example}
\label{ex:care} This example demonstrates that some care must be
taken when considering discrete and continuous variables as
mixed-pair random variables. Consider the following continuous
random variable $Y_1$ that is uniformly distributed in the
interval $[0,2]$. Now, consider the mixed random variable $V_2$
that takes the value $2$ with probability $\frac 12$ and takes a
value uniformly distributed in the interval $[0,1]$ with
probability $\frac{1}{2}$.

Clearly, there is a mapping that allows us to create $V_2$ from
$Y_1$ by just mapping $Y_1 \in (1,2]$ to the value $V_2 = 2$ and
by setting $Y_1 = V_2$ when $Y_1 \in [0,1]$. However, given
$V_2=2$ we are not able to reconstruct $Y_1$ exactly. Therefore,
intuitively one expects that $\H(Y_1) > \H(V_2)$.

However, if you use the respective injections, say $Y_1 \to Z_1$
and $V_2 \to Z_2$, to the space of mixed-pairs of random
variables, we can see that
\[ \H(Y_1) = \H(Z_1) = \log 2 = \H(Z_2). \]
This shows that if we think of $\H(Z_2)$ as the entropy of the
mixed random variable $V_2$ we get an intuitively paradoxical
result where $\H(Y_1) = \H(V_2)$ where in reality one would expect
$\H(Y_1) > \H(V_2)$.

The careful reader will be quick to point out that the injection
from $V_2$ to $Z_2$ introduces a new continuous variable,
$Y_{22}$, associated with the discrete value of $2$, as well as a
discrete value $x_0$ associated with the continuous part of $V_2$.
Indeed the "new" random variable $Y_{22}$ allows us to precisely
reconstruct $Y_1$ from $Z_2$ and thus complete the inverse mapping
of the bijection.
\end{example}
\begin{remark}
The examples show that when one has mappings involving various
types of random variables and one wishes to use bijections to
compare their entropies; one can perform this comparison as long
as the random variables are thought of as mixed-pairs.
\end{remark}

\subsection{Vector of Mixed-Pair Random Variables}

Now, we derive sufficient conditions for entropy preservation
under bijection between vectors of mixed-pair variables. To this
end, let $Z_1 = (Z_1^1,\dots,Z_d^1)$ and $Z_2 =
(Z_1^2,\dots,Z_d^2)$ be two vectors of mixed-pair random variables
with their support on $\bbS_1 \times \bbR^d$ and $\bbS_2 \times
\bbR^d$ respectively. (Here $\bbS_1, \bbS_2$ are countable subsets
of $\bbR^d$.) Let $F: \bbS_1 \times \bbR^d \to \bbS_2 \times
\bbR^d$ be a continuous and differentiable bijection that induces
$Z_2$ by its application on $Z_1$.

As before, let the projections of $F$ be $F_d : \bbS_1 \times
\bbR^d \to \bbS_2$ and $F_c: \bbS_1 \times \bbR^d \to \bbR^d$. We
consider situation where $F_c$ is differentiable. Let $g_i(\ty),
\ty \in \bbR^d$ for $\tx_i \in \bbS_1$ and $h_j(\ty), \ty \in
\bbR^d$ for $\tw_j \in \bbS_2$ be density functions as defined
before. Let $(\tx_i, \ty^1) \in \bbS_1 \times \bbR^d$ be mapped to
$(\tw_j, \ty^2) \in \bbS_2 \times \bbR^d$. Then, consider $d\times
d$ Jacobian
$$ J(\tx_i, \ty^1) \equiv \left[ \frac{\partial y^2_k}{\partial y^1_l}\right]_{1\leq k, l\leq d},$$
where we have used notation $\ty^1=(y^1_1,\dots,y^1_d)$ and
$\ty^2=(y^2_1,\dots,y^2_d)$. Now, similar to Lemma \ref{le:bijpre}
we obtain the following entropy preservation for bijection between
vector of mixed-pair random variables.
\begin{lemma}
\label{le:bijprevec} If for all $(\tx_i, \ty^1) \in \bbS_1 \times
\bbR^d$,
$$ \left|{\sf det}(J(\tx_i, \ty^1)) \right| = 1, $$
then $\H(Z^1) = \H(Z^2)$. Here ${\sf det}(J)$ denotes the
determinant of matrix $J$.
\end{lemma}
\begin{proof}
The main ingredients for the proof of Lemma \ref{le:bijpre} for
the scalar case were the equalities \eqref{eq:chofmeasure} and
\eqref{eq:chofmeasure1}. For a vector of mixed-pair variable we
will obtain the following equivalent equalities: For change of
$d\ty^1$ at $(\tx_i, \ty^1)$, let $d\ty^2$ be induced change at
$(\tx_j, \ty^2)$. Let ${\sf vol}(d\ty)$ denote the volume of $d$
dimensional rectangular region with sides given by components of
$d\ty$ in $\bbR^d$. Then,
\begin{equation}
\label{eq:chofmeasure-d}
\begin{split}
& g_i(\ty^1) {\sf vol}(d\ty^1)= h_j(\ty^2) {\sf vol}(d\ty^2).
\end{split}
\end{equation}
Further, at $(\tx_i, \ty^1)$,
\begin{equation}
\label{eq:chofmeasure1-d}
\begin{split}
& {\sf vol}(d\ty^2)=  \left|{\sf det}(J(\tx_i, y^1)) \right|{\sf
vol}(d\ty^1).
\end{split}
\end{equation}
Using exactly the same argument that is used in
\eqref{eq:bijpreserve} (replacing $dy_k$ by ${\sf vol}(d\ty^k)$,
$k=1,2$), we obtain the desired result.  This completes the proof
of Lemma \ref{le:bijprevec}.
\end{proof}

\begin{remark}
Essentially, for every discrete choice $X$, if the mapping between
the continuous vectors has  one, then the bijection preserves
entropy.
\end{remark}

\medskip
%

%
\section{Entropy Rate of Continuous Time Markov Chains}\label{se:sec3}

A continuous time Markov chain is composed of the point process
that characterizes the time of transitions of the states as well
as the discrete states between which the transition happens.
Specifically, let $x_i \in \bbR$ denote the time of $i^{th}$
transition or jump with $ i \in \bbZ$. Let $V_i \in \bbS$ denote
the state of the Markov chain after the jump at time $x_i$, where
$\bbS$ be some countable state space. For simplicity, we assume
$\bbS = \bbN$. Let transition probabilities be $p_{k\ell} =
\bbP(V_i=\ell | V_{i-1}=k), k,\ell \in \bbN$ for all $i$.

We recall that the entropy rate of a point process $\cP$ was
defined in section 13.5 of \cite{dav88} according to the
following: ``Observation of process conveys information of two
kinds: the actual number of points observed and the location of
these points given their number.'' This led them to define the
entropy of a realization $\{x_1, ..., x_N\}$ as \[ \H(N) +
\H(x_1,...,x_N|N) \] The entropy rate of the point process $\cP$
is defined as follows: let $N(T)$ be the number of points arrived
in time interval $(0,T]$ and the instances be
$\tx(T)=(x_1,\dots,x_{N(T)})$. Then, the entropy rate of the
process is
$$ \H_{ER}(\cP) = \lim_{T\to\infty} \frac{1}{T} \left[\H(N(T)) +
\H(\tx(T)|N(T))\right],$$ if the above limit exists.

We extend the above definition to the case of Markov chain in a
natural fashion. Observation of a continuous time Markov chain
 over a time interval $(0,T]$ conveys information of
three types: the number of points/jumps of the chain in the
interval, the location of the points given the number as well as
the value of the chain after each jump. Treating each random
variable as a mixed-pair allows us to consider all the random
variables in a single vector.

As before, let $N(T)$ denote the number of points in an interval
$(0,T]$. Let $\tx(T) = (x_1,...,x_{N(T)})$,
$\TV(T)=(V_0,V_1,...,V_{N(T)})$ denote the locations of the jumps
as well as the values of the chain after the jumps. This leads us
to define the entropy of the process during the interval $(0,T]$
as
\begin{equation}
\label{eq:entrate-ctmc} \H_{(0,T]} = \H(N(T),\TV(T),\tx(T)).
\end{equation}
Observe that the $(N(T), \TV(T), \tx(T))$ is a random vector of
mixed-pair variables.

For a single state Markov chain the above entropy is the same as
that of the point process determine the jump/transition times.
Similar to the development for point processes, we define the
entropy rate of the Markov chain as
\[ \H_{ER} = \lim_{T \to \infty} \frac{\H_{(0,T]}}{T}, ~~\mbox{if it exists.}
\]
\begin{proposition}\label{prop:1}
Consider a Markov chain with underlying Point process being
Poisson of rate $\lambda$, its stationary distribution being $\pi
= (\pi(i))$ with transition probability matrix $P=[p_{ij}]$. Then,
its entropy rate is well-defined and
$$ \H_{ER} = \lambda(1-\log \lambda) + \lambda {\sf{H}_{MC}},$$
where ${\sf{H}_{MC}} = - \sum_i \pi(i) \sum_{j} p_{ij} \log
p_{ij}$.
\end{proposition}
\begin{proof}
For Markov Chain as described in the statement of proposition, we
wish to establish that
$$ \lim_{T\to\infty} \frac{\H_{(0,T]}}{T} = \H_{ER},$$
as defined above. Now
\begin{equation*}
\begin{split}
\H_{(0,T]}& = \H(\tx(T), N(T), \TV(T)) \\
& = \H(\tx(T),N(T)) + \H(\TV(T)|N(T),\tx(T)).
\end{split}
\end{equation*}

Consider the term on the right hand side of the above equality.
This corresponds to the points of a Poisson process of rate
$\lambda$. It is well-known (cf. equation (13.5.10), pg. 565
\cite{dav88}) that
\begin{equation}
\label{eq:poien} \lim_{T\to\infty} \frac{1}{T}\H(\tx(T),N(T)) =
\lambda(1-\log \lambda).
\end{equation}

Now consider the term $\H(\TV(T)|\tx(T),N(T))$. Since $\TV(T)$ is
independent of $\tx(T)$, we get from the definition of conditional
entropy that
\begin{equation}
\label{eq:poiindep} \H(\TV(T)|\tx(T),N(T)) = \H(\TV(T)|N(T)).
\end{equation}
One can evaluate $\H(\TV(T)|N(T))$ as follows,
\[ \H(\TV(T)|N(T)) = \sum_k p_k \H(V_0,\dots,V_k), \]
where $p_k$ is the probability that $N(T) = k$. The sequence of
states $V_0,\dots,V_k$ can be thought of as sequence of states of
a discrete time Markov chain with transition matrix $P$. For a
Markov chain, with stationary distribution $\pi$ (i.e. $P\pi =
\pi$), it is well-known that
\begin{equation*}
\begin{split}
\lim_{k\to\infty} \frac{1}{k}\H(V_0,\dots,V_k) & = - \sum_i \pi(i) \sum_{j} p_{ij} \log p_{ij}  \\
& = {\sf{H}_{MC}}.
\end{split}
\end{equation*}
Thus, for any $\epsilon > 0$, there exists $k(\epsilon)$ large
enough such that for $k > k(\epsilon)$
\[ \left|\frac{1}{k}\H(V_0,\dots,V_k) - {\sf{H}_{MC}}\right| < \epsilon.\]
For $T$ large enough, using tail-probability estimates of Poisson
variable it can be shown that
\[ \bbP\left(N(T) \leq k(\epsilon)\right) \leq \exp\left(-\frac{\lambda
T}{8}\right).\] Putting these together, we obtain that for given
$\epsilon$ there exists $T(\epsilon)$ large enough such that for
$T \geq T(\epsilon)$
\begin{eqnarray*}
\frac{\H(\TV(T)|N(T))}{T} & = &
\frac{1}{T} \left(\sum_k k p_k \frac{\H(V_0,\dots,V_k)}{k}\right)  \\
& = &  \frac{\sum_{k \geq k(\epsilon)} k p_k ({\sf{H}_{MC}} \pm
\epsilon) +
O(k(\epsilon))}{T}  \\
& = & ({\sf{H}_{MC}} \pm \epsilon) \frac{\lambda T + O(k(\epsilon))}{T} \\
& = & \lambda {\sf{H}_{MC}} \pm 2\epsilon.
\end{eqnarray*}
That is
\[\lim_{T\to\infty} \frac{\H(\TV(T)|N(T))}{T} = \lambda {\sf{H}_{MC}}. \]

Combining \eqref{eq:poien}, \eqref{eq:poiindep} and the above
equation we complete the proof of the Proposition \ref{prop:1}.
\end{proof}

\medskip

\begin{fact}[cf. Ch. 13.5 \cite{dav88}]
\label{fa:factpoimax} Consider the set of stationary ergodic point
processes with mean rate $\lambda$. Then the entropy of this
collection is maximized by a Poisson Process with rate $\lambda$.
That is, if $\cP$ is a stationary ergodic point process with rate
$\lambda$ then
\[ \H_{ER}(\cP) \leq \lambda(1-\log \lambda). \]
\end{fact}

\section{Application}\label{se:sec4}

\subsection{Computation of continuous entropies}

In this section we show how our previous results aid in the
computation of traditional continuous time entropies. Let
$(X_1,X_2)$ be two i.i.d.\ random variables whose distributions
satisfy the conditions required of it to be a good-mixed pair. Let
$(Y_1,Y_2)$, $Y_1 < Y_2$ be the ordering of $(X_1,X_2)$. Then the
following holds:

\begin{lemma}
\label{le:orderrv} $h(Y_1,Y_2) = h(X_1,X_2) - \log 2$.
\end{lemma}
\begin{proof}
Let $\mathbb{I}$ represent the indicator function such that
$\mathbb{I}=0$ implies $X_1 = Y_1$, and $\mathbb{I}=1$ implies $X_1
= Y_2$. Clearly $\mathbb{I}$ is independent of $Y_1, Y_2$ and
probability of $\mathbb{I}=1$ is $\frac{1}{2}$. It is further easy
to see that $(\mathbb{I},Y_1,Y_2) \leftrightarrow (X_1,X_2)$ with
the corresponding Jacobians evaluating to $1$. Thus, viewed as
mixed-pairs we can equate the entropies, yielding
\[ \log 2 + h(Y_1,Y_2) = h(X_1, X_2).\]
This can be of course be shown using traditional methods but the
proof here is an illustration of how our results can be used to
obtain such results in an easier fashion.
\end{proof}

In a similar fashion if $(Y_1,...,Y_n)$ is an ordering, in
increasing order, of the i.i.d.\ random variables $(X_1,...,X_n)$,
then
$$ h(Y_1,..,Y_n) = h(X_1,....,X_n) - \log (n!) $$

\subsection{Poisson Splitting via Entropy Preservation}\label{se:poissplit}

In this section, we use the sufficient conditions developed in
Lemma \ref{le:bijprevec} to obtain proof of the following
property.

\begin{lemma}\label{le:poissplit}
Consider a Poisson process, $\cP$, of rate $\lambda$. Split the
process $\cP$ into two baby-processes $\cP_1$ and $\cP_2$ as
follows: for each point of $\cP$, toss an independent coin of bias
$p$. Assign the point to $\cP_1$ if coin turns up head, else
assign it to $\cP_2$. Then, the baby-processes $\cP_1$ and $\cP_2$
have the same entropy rate as Poisson processes of rates $\lambda
p $ and $\lambda (1-p)$ respectively.
\end{lemma}

\begin{proof}
Consider a Poisson Process, $\cP$, of rate $\lambda$ in the
interval $[0,T]$. Let $N(T)$ be the number of points in this
interval and let $\ta(T) = \{a_1,...,a_{N(T)}\}$ be their
locations. Further, let $\tC(T) = \{C_1,...,C_{N(T)}\}$ be the
outcomes of the coin-tosses and $M(T)$ denote the number of heads
among them. Denote $\tbr(T) = \{R_1,....,R_{M(T)}\}$, $\tbb(T) =
\{B_1,....,B_{N(T)-M(T)}\}$ as the locations of the baby-processes
$\cP_1, \cP_2$ respectively.

It is easy to see that the following bijection holds:
\begin{equation}
\begin{split}
& \{ \ta(T), \tC(T), N(T), M(T) \} \bij \\
& \qquad \{\tbr(T), \tbb(T), N(T)-M(T), M(T). \}
\end{split}
\end{equation}
\normalsize Given the outcomes of the coin-tosses $\tC(T)$,
$\{\tbr(T),\tbb(T)\}$ is a permutation of $\ta(T)$. Hence, the
Jacobian corresponding to any realization of
$\{\tC(T),N(T),M(T)\}$ that maps $\ta(T)$ to $\{\tbr(T),\tbb(T)\}$
is a permutation matrix, i.e  determinant is $\pm 1$.

Therefore, Lemma \ref{le:bijprevec} implies that
\begin{equation}
\label{eq:ps2}
\begin{split}
&\H(\ta(T), \tC(T), N(T), M(T))      \\
&\quad = \H(\tbr(T), \tbb(T), N(T)-M(T), M(T))\\
&\quad \leq \H(\tbb(T), N(T)-M(T)) + \H(\tbr(T), M(T)).
\end{split}
\end{equation}
\normalsize $M(T)$ is completely determined by $\tC(T)$ and it is
easy to deduce from the definitions that
\[ \H(M(T)|\ta(T),\tC(T),N(T)) = 0.\]
Hence
\begin{equation}
\label{eq:ps2a}
\begin{split}
&  \H(\ta(T), \tC(T), N(T), M(T))   \\
&\quad =  \H(\ta(T), \tC(T), N(T)) + \H(M(T)|\ta(T), \tC(T), N(T)) \\\
&\quad =  \H(\ta(T), \tC(T), N(T)).
\end{split}
\end{equation}
\normalsize Since the outcome of the coin-tosses along with their
locations form a continuous time Markov chain, using Proposition
\ref{prop:1} we can see that
\begin{equation}
\label{eq:ps3}
\begin{split}
& \lim_{T \to \infty}\frac{1}{T} \H(\ta(T), \tC(T), N(T), M(T))   \\
&\quad = \lim_{T \to \infty} \frac{1}{T} \H(\ta(T), \tC(T), N(T))  \\\
&\quad = \lambda(1-\log \lambda) - \lambda(p\log p + (1-p) \log
(1-p)) \\
&\quad = \lambda p (1-\log \lambda p) + \lambda (1-p) (1-\log
\lambda (1-p)).
\end{split}
\end{equation}
\normalsize

It is well known that $\cP_1, \cP_2$ are stationary ergodic
processes of rates $\lambda p, \lambda (1-p)$ respectively. Hence
from Fact \ref{fa:factpoimax} we have
\begin{equation}
\label{eq:ps3b}
\begin{split}
& \lim_{T \to \infty}\frac{1}{T} \H(\tbr(T),M(T)) \leq \lambda p
(1-\log \lambda p),\\
& \lim_{T \to \infty}\frac{1}{T} \H(\tbb(T), N(T)-M(T)) \leq
\lambda (1-p) (1-\log \lambda (1-p)).
\end{split}
\end{equation}
\normalsize

Combining equations \eqref{eq:ps2}, \eqref{eq:ps3},
\eqref{eq:ps3b} we can obtain
\begin{equation}
\label{eq:ps4}
\begin{split}
& \lim_{T \to \infty}\frac{1}{T} \H(\tbr(T),M(T)) = \lambda p
(1-\log \lambda p),\\
& \lim_{T \to \infty}\frac{1}{T} \H(\tbb(T), N(T)-M(T))  = \lambda
(1-p) (1-\log \lambda (1-p)).
\end{split}
\end{equation}
Thus, the entropy rates of processes $\cP_1$ and $\cP_2$ are the
same as that of Poisson processes of rates $\lambda p$ and
$\lambda (1-p)$ respectively. This completes the proof of Lemma
\ref{le:poissplit}.
\end{proof}


\section{Conclusions}\label{se:sec5}

This paper deals with notions of entropy for random variables that
are mixed-pair, i.e. pair of discrete and continuous random
variables. Our definition of entropy is a natural extension of the
known discrete and differential entropy. Situations where both
continuous and discrete variables arise are common in the analysis
of randomized algorithms that are often employed in networks of
queues, load balancing systems, etc. We hope that the techniques
developed here will be very useful for the analysis of such
systems and for computing entropy rates for the processes
encountered in these systems.

\bibliographystyle{amsalpha}
\bibliography{mybiblio}

\section*{Appendix}

\subsection{Proof of Lemma \ref{le:gmpcond}}

We wish to establish that the conditions of Lemma \ref{le:gmpcond}
guarantee that
\begin{equation}
\label{eq:gmpsuff} \sum_i \int g_i(y) | \log g_i(y) | dy < \infty.
\end{equation}

Let $(a)_+ = \max(a, 0)$ and $(a)_{-} = \min(a, 0)$ for $a \in
\bbR$. Then, $$a = a_+ + a_-, \quad \mbox{and} \quad |a| = a_+ -
a_-.$$ By definition $g_i(y) \geq 0$. Observe that
\[ | \log g_i(y) | = 2 ( \log g_i(y) )_+ - \log g_i(y). \]

Therefore to guarantee \eqref{eq:gmpsuff} it suffices to show the
following two conditions:
\begin{eqnarray}
\label{eq:gmpsuff1} &&\sum_i \int_\bbR g_i(y) ( \log g_i(y) )_+ dy
<
\infty, \\
&& \sum_i \left| \int_\bbR g_i(y) \log g_i(y) dy \right| < \infty.
\label{eq:gmpsuff2}
\end{eqnarray}

The next two lemmas show that equations \eqref{eq:gmpsuff1} and
\eqref{eq:gmpsuff2} are satisfied and hence completes the proof of
Lemma \ref{le:gmpcond}.

\begin{lemma}
\label{le:gmpprelim1} Let $Y$ be a continuous random variable with
a density function $g(y)$ such that for some $\delta > 0$
$$\int_{\bbR} g(y)^{1+\delta} dy < \infty.$$
Further if $g(y)$ can be written as sum of non-negative functions
$g_i(y)$, the
\[ \sum_i \int_\bbR g_i(y) (\log g_i(y))_+ dy < \infty. \]
\end{lemma}

\medskip

\begin{proof}
For given $\delta$, there exists finite $B_\delta > 1$ such that
for $x \geq B_\delta$, $\log x \leq x^{\delta}$. Using this, we
obtain
\begin{equation}
\begin{split}
& \int_\bbR g_i(y) (\log g_i(y))_+ dy  =  \int_{g_i(y) \geq 1} g_i(y) \log g_i(y) dy \\
 & \quad =  \int_{1\leq g_i(y) < B_{\delta}} g_i(y) \log g_i(y) dy +  \int_{B_{\delta}
 \leq g_i(y)} g_i(y) \log g_i(y) dy \\
& \quad\leq  \log B_{\delta} \int_{\bbR} g_i(y) dy + \int_{\bbR} g_i(y)^{1+\delta} dy \;\\
& \quad = p_i \log B_{\delta}  + \int_{\bbR} g_i(y)^{1+\delta} dy.
\label{eq:equation0}
\end{split}
\end{equation}

Therefore,
\begin{equation*}
\begin{split}
& \sum_i \int_\bbR g_i(y) (\log g_i(y))_+ dy \leq \sum_i \left(p_i
\log B_{\delta}  + \int_{\bbR}
g_i(y)^{1+\delta} dy \right) \\
& \quad \stackrel{(a)}{=} \log B_{\delta} +  \int_{\bbR} \sum_i
g_i(y)^{1+\delta} dy \\
&\quad \stackrel{(b)}{\leq} \log B_{\delta} + \int_{\bbR}
g(y)^{1+\delta} dy  < \infty.
\end{split}
\end{equation*}
In $(a)$ we use the fact that $g_i(y)$ is positive to interchange
the sum and the integral. In $(b)$, we again use the fact that
$g_i(y) \geq 0$ to bound $\sum_i g_i(y)^{1+\delta}$ with
$\left(\sum_i g_i(y)\right)^{1+\delta}$.

\end{proof}

\begin{lemma}
\label{le:ap2} In addition to the hypothesis of $Y$ in Lemma
\ref{le:gmpprelim1} assume that $Y$ has a finite $\epsilon$ moment
for some $\epsilon
> 0$. Then the following holds:
\[ \sum_i \left| \int g_i(y) \log g_i(y) dy \right| < \infty. \]
\end{lemma}

\begin{proof}
Let for some $\epsilon > 0$,
\[ M_\epsilon = \int_\bbR |y|^\epsilon g(y) \; dy < \infty. \]

Note that for any $\epsilon > 0$, there is a constant $C_\epsilon
> 0$, such that  $\int_{\bbR}C_\epsilon  e^{-|y|^\epsilon} dy \; =
1$. Further, observe that the density $\tilde{g}_i(y) =
g_i(y)/p_i$ is absolutely continuous w.r.t. the density $f(y)
(\stackrel{\triangle}{=} C_\epsilon e^{-|y|^\epsilon})$. Thus from
the fact that the Kullback-Liebler distance $D(\tilde{g}_i||f)$ is
non-negative we have
\begin{equation*}
\begin{split}
 0 &\leq  \int_\bbR g_i(y) \log \frac{g_i(y)}{p_i f(y)} \; dy
 =  \int_\bbR g_i(y) \log \frac{g_i(y)}{p_i C_\epsilon
 e^{-|y|^\epsilon}} \; dy\\
 &=  \int_\bbR g_i(y) \log g_i(y) \; dy - p_i \log p_i  -  p_i \log C_\epsilon + \int_\bbR
 |y|^\epsilon g_i(y) dy. \;
 \end{split}
 \end{equation*}

Therefore
\begin{equation}
\begin{split}
\label{eq:equation1} &- \int_\bbR g_i(y) \log g_i(y) \; dy \leq -
p_i \log p_i    + p_i |\log C_\epsilon| + \int_\bbR
 |y|^\epsilon g_i(y) dy.
\end{split}
\end{equation}

\medskip

From \eqref{eq:equation0} we have
\begin{equation}
\begin{split}
\label{eq:equation2} & \int_\bbR g_i(y) \log g_i(y) \; dy \leq
\int_\bbR g_i(y) (\log g_i(y))_+ dy \leq p_i \log B_{\delta}  +
\int_{\bbR} g_i(y)^{1+\delta} dy.
\end{split}
\end{equation}

\medskip
\medskip

Combining equations \eqref{eq:equation1} and \eqref{eq:equation2},
we obtain
\begin{equation}
\begin{split}
 \left| \int_\bbR g_i(y) \log g_i(y) \; dy \right|  &\leq - p_i
\log p_i    + p_i |\log C_\epsilon| + \int_\bbR
 |y|^\epsilon g_i(y)
  + p_i \log B_{\delta} \\
& \quad   + \int_{\bbR} g_i(y)^{1+\delta} dy.
\end{split}
\label{eq:equation3}
\end{equation}

\medskip
\medskip

Now using the facts
\begin{equation*}
\begin{split}
& -\sum_i p_i \log p_i < \infty,  ~~\sum_i \int_\bbR |y|^\epsilon
g_i(y) dy  = \int_\bbR |y|^\epsilon
 g(y) dy < \infty, \\
 &~\mbox{and}
  ~~\sum_i \int_{\bbR} g_i(y)^{1+\delta} dy < \int_{\bbR}
g(y)^{1+\delta} dy < \infty,
\end{split}
\end{equation*}
we obtain from \eqref{eq:equation3} that
\[ \sum_i \left| \int g_i(y) \log g_i(y) dy \right| < \infty.\]

\end{proof}

\end{document}